\documentclass[twocolumn,superscriptaddress,showpacs,10pt,prl]{revtex4}

\usepackage{graphicx}%

\DeclareGraphicsExtensions{.eps}

\newcommand{\re}{{\text{Re}}}
\newcommand{\im}{{\text{Im}}}

\begin{document}

\title{High Frequency Conductivity in the Quantum Hall Regime}

\author{F.~Hohls}
\email{hohls@nano.uni-hannover.de}
\author{U.~Zeitler}%
\author{R.~J.~Haug}
\affiliation{%
Institut f\"ur Festk\"orperphysik,
Universit\"at Hannover, Appelstr. 2, 30167 Hannover, Germany
}%

\date{\today}

\begin{abstract}

We have measured the complex conductivity
$\sigma_{xx}$ of a two-dimensional electron system in the quantum Hall 
regime up to frequencies of 6~GHz at electron temperatures
below 100 mK. Using both its imaginary and real part 
we show that $\sigma_{xx}$ can be scaled to a single function
for different frequencies and for all 
investigated transitions between plateaus in the quantum Hall effect.
Additionally, the conductivity in the variable-range hopping regime 
is used for a direct evaluation  of the localization length $\xi$.
Even for large filing factor distances $\delta \nu$
from the critical point we find $\xi \propto \delta \nu^{-\gamma}$
with a scaling exponent $\gamma=2.3$.

\end{abstract}


\maketitle

\newlength{\plotwidth}          
\setlength{\plotwidth}{0.9\linewidth} 

It is widely accepted that 
the understanding of the integer quantum Hall effect (QHE) is closely related
to a disorder driven localization-delocalization transition occurring
in two-dimensional electron systems (2DES) in high magnetic fields \cite{review}. 
 Many experimental and theoretical works
approve the interpretation of the transition between adjacent QHE plateaus
as a quantum critical phase transition. It is governed by a 
diverging localization length 
$\xi \propto |E-E_c|^{-\gamma}$ which scales with the
distance of the energy $E$ from the critical 
energy $E_c$ in the center of a Landau band. 
The exponent $\gamma\approx 2.3$ is believed to be 
a universal quantity independent of
disorder. For finite systems with effective
size $L_{\text{eff}}$ theory predicts that the
conductivities $\sigma_{\alpha\beta}$ follow scaling
functions $\sigma_{\alpha\beta}=
f_{\alpha\beta}\left( L_{\text{eff}} / \xi(E) \right)$ 
resulting in a finite width 
$\Delta E \propto L_{\text{eff}}^{-1/\gamma}$ of the transition region.
The effective system size $L_{\text{eff}}$ is determined by the physical
sample size, the electron temperature $T$ or the frequency $f$. 

The most common test of scaling uses an analysis of the
temperature or frequency dependence of the conductivity peak width
in the QHE plateau transition.
However, lacking an exact expression
for \mbox{$L_{\text{eff}}(T,f)$}, this method does not allow to access 
the scaling behavior of the localization length directly.

An alternative approach to 
scaling was proposed by Polyakov and Shklovskii \cite{polyakov93}.
Using the fact that the conductivity $\sigma_{xx}$ 
in the QHE plateaus at low tem\-pe\-ra\-tures is dominated by 
variable-range hopping (VRH) \cite{vrh1}, 
they analyzed the temperature, current, and frequency dependence 
of $\sigma_{xx}$ in order to 
gain direct access to the localization length $\xi$.
The temperature dependence of $\sigma_{xx}$ in the VRH regime
was also investigated experimentally \cite{vrh2}. 
However, due to an unknown theoretical prefactor, 
$\xi$ could only be estimated from these experiments.

In contrast, the frequency driven variable-range hopping 
conductivity $\sigma_{xx}(f)$
(in the limit \mbox{$\sigma_{xx}(f)\gg \sigma_{xx}(0)$}) is given
by \cite{polyakov93}
\begin{equation}
  \label{hoppingeq}
  \re \,\sigma_{xx}(\omega) = \frac{2\pi}{3}\epsilon\epsilon_0\xi\omega,
\end{equation}
linearly depending on both frequency \mbox{$f=\omega/2\pi$}
and localization length $\xi$ with no unknown  prefactors.

Here we report on measurements of the 
complex conductivity $\sigma_{xx}$ up to frequencies $f = 6$~GHz at low 
temperatures down to below 100 mK. We will show that 
$\im(\sigma_{xx})$ can be scaled with a single-parameter function
to $\re(\sigma_{xx})$, independent of temperature, frequency, and 
filling factor. Secondly we will use Eq. (\ref{hoppingeq})
to directly measure the localization length $\xi$ deep into the
variable-range hopping regime for several QHE plateau transitions.
Its filling factor dependence follows a scaling behavior 
\mbox{$\xi\propto |\nu-\nu_c|^{-\gamma}$}
with \mbox{$\gamma=2.3$}
up to large distances \mbox{$|\nu-\nu_c|\geq 0.3$} from the critical 
point $\nu_c$.
In the center of the QHE plateaus $\xi$ is found to
be limited by the magnetic length, the natural length scale
of the quantum Hall state.

The two-dimensional electron system used in our experiments
was realized in an \mbox{AlGaAs/GaAs}
heterostructure grown by molecular beam epitaxy and lies 75~nm underneath the
surface. Its electron mobility and density are \mbox{$\mu=35$~m$^2$/Vs} and 
\mbox{$n=3.3\cdot 10^{15}$~m$^{-2}$}. 
The sample was patterned into Corbino geometry with
contacts fabricated by standard Ni/Au/Ge alloy annealing. 
This geometry allows
a direct two-point measurement of the longitudinal
conductivity $\sigma_{xx}$ at high frequencies.
For an ideal Corbino geometry $\sigma_{xx}$ is given by 
\mbox{$2\pi \sigma_{xx}={G}\ln\left({r_2}/{r_1}\right)$}.
\mbox{$G=I/U$} is the two-point conductance with current $I$ and 
voltage $U$ measured at the same contacts.
The sample dimensions of our Corbino ring 
are \mbox{$r_2=820\,\mu$m} for the outer and \mbox{$r_1=800\,\mu$m}
for the inner radius. 

The conductance measurement at high frequencies is realized by a reflection
measurement setup: The sample acts as load of a high frequency coaxial 
line with a characteristic impedance of \mbox{$Z_0=50\,\Omega$}. 
A load impedance \mbox{$Z=1/G$} deviating
from $Z_0$ leads to reflection of an incident wave at the load
with a complex reflection
coefficient \mbox{$\mathcal{R}_{S}=(Z-Z_0)/(Z+Z_0)$}. The
total reflection coefficient $\mathcal{R}$ of the loaded 
line is made up of the sample reflection $\mathcal{R}_S$
and of properties of the line itself such as
phase shifts, losses and reflections at interconnections.
A careful calibration of these frequency dependent contributions
of the line allows the extraction of \mbox{$\mathcal{R}_S(f)$} 
and therefore of 
the complex sample conductivity $\sigma_{xx}(f)$
from the direct accessible quantity \mbox{$\mathcal{R}(f)$}.
The sample dimensions were chosen as compromise between sensitivity,
which is largest for a sample impedance $Z$ close to \mbox{$Z_0=50\,\Omega$},
and the avoidance of size effects for too small ring widths \cite{koch91}.

Sample and coaxial line where fitted into a dilution refrigerator with base
temperature \mbox{$T_S < 50$~mK} 
using a multi step thermal sinking of the line.
The sample is situated in the center of a super-conducting solenoid capable of
producing magnetic fields up to \mbox{$B=15\,$T}. The reflection 
$\mathcal{R}$ of 
the coaxial line terminated with the sample is
measured with a network analyzer with a frequency range \mbox{$f=100$~kHz}
to 6~GHz
using power levels \mbox{$P\leq -75$~dBm} which were checked not to 
influence the measured conductivity. The analyzer was used in
continuous wave mode with fixed frequency  while stepping the magnetic field.

\begin{figure}[tb]
  \begin{center}
   \resizebox{\plotwidth}{!}{\rotatebox{0}{\includegraphics{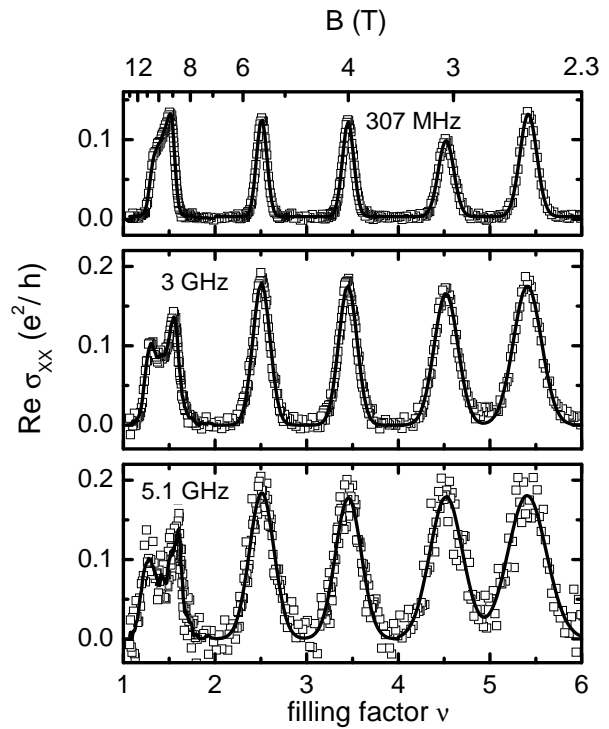}}}
  \end{center}
  \caption{Real part of the conductivity at different frequencies. The squares
       are the raw data points, the line is derived by Gaussian fits for 
       \mbox{$\nu>2$} and simple smoothing for \mbox{$\nu<1$}.}
  \label{data}
\end{figure}

Figure \ref{data} shows the real part $\re(\sigma_{xx})$
of the measured conductivity as function
of the filling factor \mbox{$\nu=n/\!\left(eB/h\right)$}.
It shows well
pronounced Shubnikov-de Haas oscillations with 
zero conductivity at integer filling factors and maxima near half integer
$\nu$. Every peak corresponds to the transition between adjacent 
quantum Hall states with a critical point $\nu_c$ at maximum conductivity. 
The transition \mbox{$\nu=1\rightarrow2$} 
within the lowest Landau level deviates
from the other transitions by form and amplitude of the peak and shows 
a shoulder possibly originating from the presence of 
an impurity band \cite{kuchar00}.

The other transitions are very symmetric and can be fitted
by Gaussians.
This fit is used to derive the position of the critical filling factors
$\nu_c$ and the critical conductivities
$\sigma_c$. 
For \mbox{$f > 2$~GHz} and \mbox{$\nu > 2$}
its value \mbox{$\sigma_c\approx 0.17\,e^2/h$} 
is the same for all transitions and independent of frequency as expected by
scaling theory, but deviates from the proposed universal value 
\mbox{$0.5\,e^2/h$} \cite{sigma}. 
A possible explanation for
this discrepancy was given by Ruzin \textsl{et al.} \cite{ruzin96} 
who proposed fluctuations in the carrier density
as cause of non-universal critical conductivity values.

\begin{figure}[tb]
  \begin{center}
  \resizebox{\plotwidth}{!}{\rotatebox{0}{\includegraphics{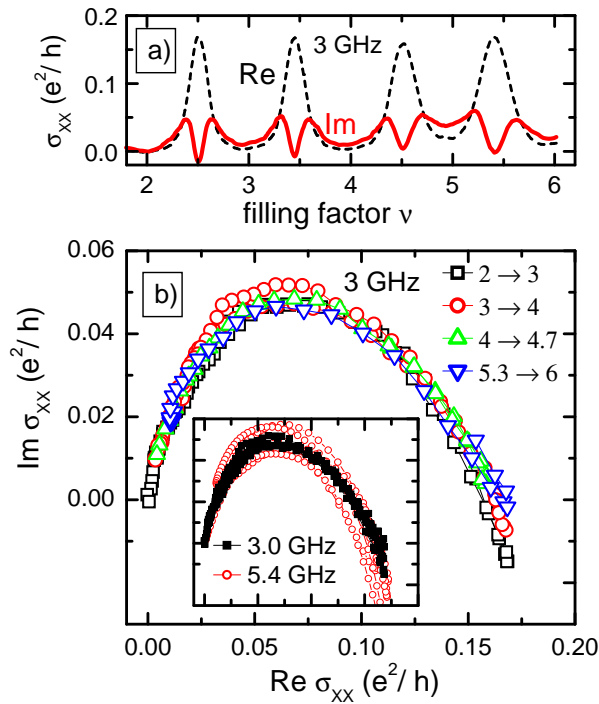}}}
  \end{center}
  \caption{a) Real (broken line) and imaginary (full line) part of
        $\sigma_{xx}$ versus filling factor for \mbox{$f=3$~GHz}.
        b) Plot of $\im(\sigma_{xx})$ vs. $\re(\sigma_{xx})$ for 
        \mbox{$f=3$~GHz}. 
        The four curves for different plateau transitions, 
        marked with different symbols, coincide.
        The data around the not fully resolved plateau $\nu=5$
        are omitted.
        Inset: The same plot for two different  frequencies 
        (\mbox{$f=3$~GHz} and \mbox{$f=5.4$~GHz}), each 
        curve including all four transitions in the range $\nu=2-6$.
        Again the curves collapse to a single function.
        }
  \label{path}
\end{figure}

Previous experiments by Engel \textsl{et al.} \cite{engel93} 
and Balaban \textsl{et al.} \cite{balaban98} 
were restricted to a measurement  of the real 
part of $\sigma_{xx}$ whereas our technique gives access to both the 
real and imaginary part. This allows an additional test
of universal frequency scaling. 
As an example we show the real and the imaginary part of the 
complex conductivity
$\sigma_{xx}$ for \mbox{$f=3$~GHz} in Fig.~\ref{path}a.
$\im(\sigma_{xx})$ and $\re(\sigma_{xx})$ display a similar symmetry around
the critical points, which are marked by the maxima of the conductivity.
In the QHE plateau centers between two critical points $\sigma_{xx}$
tends to zero (except for $\nu=5$ where spin splitting is no longer 
fully resolved). 
Approaching the critical points both $\im(\sigma_{xx})$
and $\re(\sigma_{xx})$ start to rise to positive value.
While to our knowledge there is no published theoretical 
prediction of the imaginary part of the hopping conductivity of interacting 
electrons in high magnetic fields this
behavior agrees qualitatively with the theory of Efros \cite{efros85} for
low magnetic fields. He calculated 
$\im(\sigma_{xx})\propto \ln\left(\omega_{\text{ph}}/\omega\right) 
\re(\sigma_{xx})$ 
with $\omega_{\text{ph}}$ the characteristic phonon frequency, 
which gives a linear
dependence between $\im(\sigma_{xx})$ and $\re(\sigma_{xx})$ 
at fixed frequency.  

When moving closer towards the critical point $\im(\sigma_{xx})$ starts
dropping back to a a value near zero whereas $\sigma_{xx}$ continues 
increasing up to $\sigma_{xx}^{max} = 0.17\,e^2/h$ for the sample investigated. 
This agrees with the expectation for a quasi metallic behavior
of a 2DES at the critical point. 
The crossover between the variable-range hopping regime and the metallic 
regime occurs at a conductivity 
\mbox{ $ \re(\sigma_{xx}) \approx 0.4\,\sigma^{\text{max}}_{xx} $ },
independent of frequency \mbox{($f\geq1$~GHz)} and temperature.

The extra information gained from the imaginary part allows a new test 
of the applicability of scaling theory.
The theory predicts a complex
conductivity following a scaling function 
\mbox{$\sigma_{xx}(f,\delta\nu) = g_{xx} (L_f / \xi(\delta\nu))$}~\cite{pruisken88}, 
where \mbox{$\delta\nu=\nu-\nu_c$} is
the deviation to the nearest critical point and 
\mbox{$L_f\propto f^{-z}$} is the frequency dependent dynamic length. 
Since \mbox{$\re(\sigma_{xx})$} depends monotonically
on $\delta\nu$ we can invert $\re\,\sigma_{xx}(y)$ and replace the argument
\mbox{$y=L_f/\xi(\delta\nu)$} in \mbox{$\im\,\sigma_{xx}(y)$} 
with some function \mbox{$g^{-1}(\re\,\sigma_{xx})$}.
Doing this we expect an explicit, transition and frequency independent function
\mbox{$\im (\sigma_{xx}) = \tilde{g}\left(\re (\sigma_{xx})\right)$}. 
In Fig.~\ref{path}b, where $\im (\sigma_{xx})$ vs.~$\re (\sigma_{xx})$
at \mbox{$f=3$~GHz}
is shown for different transitions, such a universal scaling is indeed
observed. 
We find a similar agreement for all frequencies
\mbox{$f\geq 1$~GHz}.
As an example the comparison of the filling factor range \mbox{$\nu=2-6$} 
is shown for frequencies of \mbox{$f=3$~GHz} and \mbox{$f=5.4$~GHz} 
in the inset of Fig.~\ref{path}b. 
Although the transition width $\Delta \nu$
for the two frequencies
differs by a factor of $1.5$ the shape in this kind of plotting of
$\im(\sigma_{xx})$ versus $\re(\sigma_{xx})$ agrees.
This insensitivity on filling factor and frequency confirms the 
scaling behavior of the sample.

\begin{figure}[tb]
  \begin{center}
  \resizebox{\plotwidth}{!}{\rotatebox{0}{\includegraphics{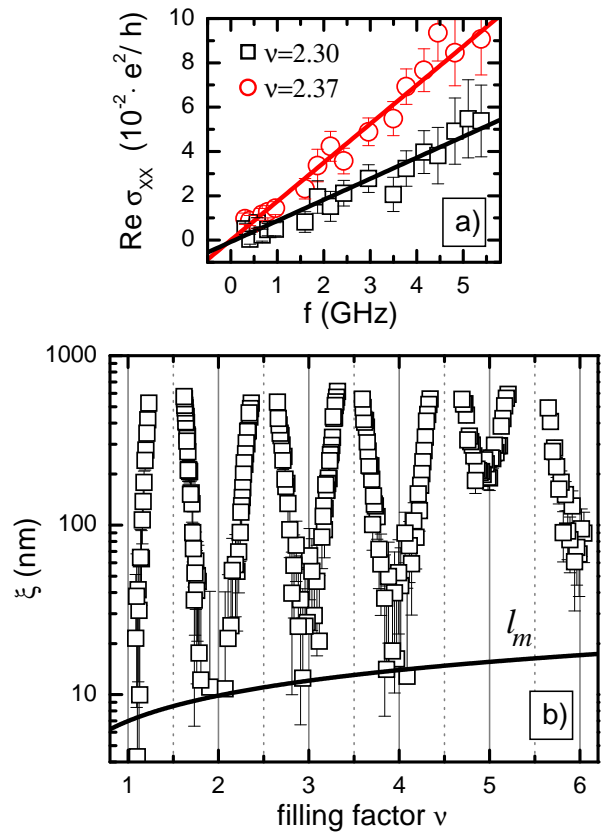}}}
  \end{center}
  \caption{a) Linear fit of the frequency dependence for two different filling
    factors. b) Localization length $\xi$ as
    function of filling factor. The error bars correspond to the 
    uncertainty of the linear fits. Around half integer filling factors
    $\sigma_{xx} (f) \gg \sigma_{xx}(0)$ is not fulfilled and thus the 
    determination of $\xi$ is not possible. The solid line denotes the
    filling factor dependence of 
    the magnetic length $l_m$.}
  \label{hopping1}
\end{figure}

Let us now turn to the second central topic of this Letter, 
namely a direct experimental access to the
localization length $\xi$.  
To test for conventional scaling behavior we have investigated
the temperature and voltage dependence of the plateau 
transition width $\delta \nu$. Both follow power-laws 
\mbox{$\Delta\nu\propto T^{\kappa}$}
and \mbox{$\Delta\nu\propto V^a$}
with exponents \mbox{$\kappa \approx 0.43$} and \mbox{$a\approx 0.22$} as found
in most previous experiments \cite{review}.
The frequency dependence of $\Delta \nu$ is also found to be 
in agreement with scaling, however with a scaling exponent
$c = 1/z\gamma \approx 0.5-0.6$ which is larger than the value
proposed by Engel \textsl{et al.} ($c=0.43$) \cite{engel93}. 
A more detailed analysis of the plateau transition width can
be found in Ref.~\onlinecite{hohls00a}.

Though the conventional analysis of the width $\Delta\nu$ of
the plateau transition  is a powerful tool for the
test of scaling, it is only an indirect
approach to the localization length. Using Eq.~(\ref{hoppingeq}) 
our high frequency measurements allow 
a direct evaluation of this length, provided that 
\mbox{$\sigma_{xx}(f)\gg \sigma_{xx}(0)$} 
for a fixed filling factor. 
Examples for linear fits of our data to Eq.~(\ref{hoppingeq})
are shown in Fig.~\ref{hopping1}a. 
Using  Eq.~(\ref{hoppingeq}) with $\epsilon\approx 12$ for GaAs the
localization length $\xi$ can be directly extracted from the
slope of these linear fits. The resulting dependence of
$\xi$ on $\nu$ is plotted in Fig.~\ref{hopping1}b.
\begin{figure}[tb]
  \begin{center}
  \resizebox{0.9\plotwidth}{!}{\rotatebox{0}{\includegraphics{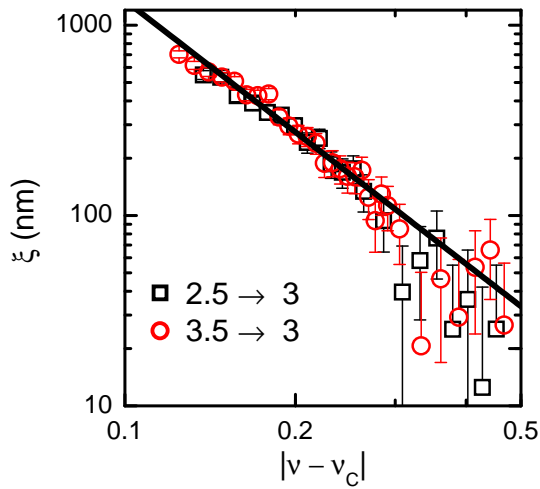}}}
  \end{center}
  \caption{Localization length $\xi$ as function
    of the filling factor difference $\delta\nu=|\nu-\nu_c|$ to the nearest
    critical point $\nu_c$ for both sides of the $\nu=3$ plateau. 
    The straight line corresponds to the expected power law
    with a universal exponent $\gamma=2.3$.}
  \label{hopping2}
\end{figure}
In the same graph the filling factor dependence of the magnetic
length $l_m=\sqrt{\hbar/eB}$ is drawn. 
Though the error bars are rather large compared to this 
length the plot shows qualitatively, that the magnetic length, which is the
shortest possible length scale of the electron system in a high magnetic field, 
sets the order  of magnitude for the 
localization length in the middle of the quantum Hall plateaus. 
The plateaus for \mbox{$\nu \geq 5$} show a larger localization length 
because the energy gaps are no longer
fully resolved at low magnetic fields.

With this direct measurement of the localization length it is
possible to test the scope of the 
power law \mbox{$\xi \propto |\nu-\nu_c|^{-\gamma}$}
when approaching the QHE plateau. 
This is shown in Fig. \ref{hopping2} for both sides of the $\nu=3$ plateau.
Data and power law with $\gamma=2.3$ agree well over the complete range.
Because of the rather large distance from the critical point
this is, on a first glimpse, rather astonishing.
However, numerical calculations
of Huckestein \textsl{et al.} \cite{huckestein90} also show a power law
\mbox{$\xi\propto |E-E_c|^{-\gamma}$} down to a
localization length of a few magnetic lengths $l_m$. 

In conclusion we measured the frequency dependence of the longitudinal
conductivity $\sigma_{xx}$ for frequencies up to 6~GHz. We were able to access
both the real part and the
imaginary part of $\sigma_{xx}$ which allowed us to perform a new test
of scaling behavior. Using the theory of Polyakov
and Shklovskii on variable-range hopping in the QHE
we were able to deduce the localization length.
Its filling factor dependence is consistent with universal
scaling behavior far into the QHE plateau.

We thank F.~Evers, B.~Huckestein, B. Kramer, F.~Kuchar, J.~Melcher, 
D.~G.~Polyakov and L.~Schweitzer
for useful discussions. The sample was grown by K.~Pierz at the 
Physikalisch-Technische Bundesanstalt in Braunschweig.

\end{document}